\documentclass[showpacs,preprintnumbers,amsmath,amssymb,eqsecnum]{revtex4}
\usepackage{graphicx}
\begin{document}
\preprint{IMAFF-RCA-06-10}
\title{Dark energy without dark energy}

\author{Pedro F. Gonz\'{a}lez-D\'{\i}az}
\affiliation{Colina de los Chopos, Centro de F\'{\i}sica ``Miguel A.
Catal\'{a}n'', Instituto de Matem\'{a}ticas y F\'{\i}sica Fundamental,\\
Consejo Superior de Investigaciones Cient\'{\i}ficas, Serrano 121,
28006 Madrid (SPAIN).}
\date{\today}
\begin{abstract}
It is proposed that the current acceleration of the universe is
not originated by the existence of a mysterious dark energy fluid
nor by the action of extra terms in the gravity Lagrangian, but
just from the sub-quantum potential associated with the CMB
particles. The resulting cosmic scenario corresponds to a benigner
phantom model which is free from the main problems of the current
phantom approaches.

\end{abstract}



\maketitle

\section{Introduction}

Besides some failed attempts to justify current acceleration of
the universe by considering dimming mechanisms in the
neighbourhood of distant supernovas (see e.g. \cite{Csaki:2005}),
essentially there are three main paradigms that have been invoked
to lend physical support to the observed speeding-up of the
universe: the cosmological constant \cite{Padmanabhan:2006}, dark
energy \cite{Turner:2001} and modified gravity
\cite{Carroll:2004}. It is well-known however that the existence
of a cosmological term poses a fundamental problem with quantum
field theory whose solution has been unsuccessfully looked for
during the last quarter of century or so \cite{Weinberg:1989}.
Then, whereas dark energy is usually implemented by introducing a
scalar field that corresponds to a fluid with negative pressure
\cite{Caldwell:1999}, the idea of modifying gravity amounts to
adding some extra terms to the Hilbert-Einstein Lagrangian
\cite{Stelle:1978} for general relativity, in both cases making
recourse to older procedures already used in inflationary
scenarios or quantum gravity. Nevertheless, all of these paradigms
are not free from remarkable shortcomings and, of course, look
quite alien to the so-called Occam Razor guiding principle; that
is, no consistent new idea based just on Einstein general
relativity and the checked contents of the universe has so far
been advanced to justify universal acceleration.

I will consider here a cosmic model where we introduce what might
be the germ of one of such ideas. In order to see how that model
works, one could however make still use of a scalar field whose
introduction would be motivated by up-grading-to field
\cite{Bagla:2003} a background set of relativistic particles,
showing then that the up-grading method becomes superfluous, so
that the actual physical ingredients of the resulting cosmic model
are just the sub-quantum characteristics that can be associated
with the original radiation particles. Our idea consists in
identifying such radiation particles with the cosmic microwave
background and what we currently call dark energy with the
radiation sub-quantum potential energy. It will be finally shown
that the resulting cosmic accelerating model describes a benigner
phantom-like cosmology which is free from the main difficulties
showed by that kind of cosmic models.

\section{The model}

Our most economical description starts with the quasi-classical
wave function for the considered particles
\begin{equation}
\Psi=R(r,t)e^{iS(r,t)/\hbar},
\end{equation}
in which $R(r,t)$ is the probability amplitude to find the
particle at position $r$ at time $t$, and $S(r,t)$ is the
corresponding classical action. Now, from the real part of the
expression resulting when applying the Klein-Gordon equation
without any potential energy term to the above wave function we
can derive the modified Hamilton-Jacobi equation
\begin{equation}
E^2 -p(v)^2 +\tilde{V}_{SQ}^2=m_0^2 ,
\end{equation}
where $E$ and $p$ are the classical energy and momentum, $m_0$ is
the rest mass, and
\begin{equation}
\tilde{V}_{SQ}=\hbar\sqrt{\frac{\nabla^2 R-\ddot{R}}{R}}
\end{equation}
is the sub-quantum potential that distinguishes the classical from
the quantum particle dynamics \cite{Bohm:1952}. Note that in the
classical limit $\hbar\rightarrow 0$ Eq. (2) becomes the classical
Hamilton-Jacobi equation. Now, a cosmic field theory could be
obtained by using the motivating up-grading method according to
which $v^2=\dot{q}^2$ and $m_0$ are respectively promoted to the
scalar field quantities $\dot{\phi}^2$ and $\tilde{V}(\phi)$
\cite{Gonz:2004}. However, it will be shown later on that in the
cosmological model we are going to build up we necessarily have
$\dot{\phi}^2=1$ and $\tilde{V}(\phi)=0$ which are conditions that
amount to convert the up-grading-to-field method into an identity
operation and the original particles into radiation particles
which thus becomes the sole physical entities, other than a
cosmological constant, entering the model.

This result can be implemented by using the following Lagrangian
density
\begin{equation}
L=-m_0\left(E(x,k)-\sqrt{1-v^2}\right),
\end{equation}
where $E(x,k)$ is the elliptic integral of the second kind
resulting from integrating the expression for the momentum derived
from Eq. (2) over the particle velocity [10], with
$x=\arcsin\sqrt{1-v^2}$ and $k=\sqrt{1-V_{SQ}^2/m_0^2}$, and
$V_{SQ}$ is the sub-quantum potential energy density. Deriving the
subsequent expressions for the energy density, $\rho$, and
pressure, $p$, in a flat Friedmann-Robertson-Walker scenario and
assuming an equation of state $p=w\rho$, with the parameter
$w\equiv w(t)$, after some rather trivial manipulations, we
finally obtain the required condition $m_0=V(\phi)=0$
\cite{GonzRozas:2006}, and
\begin{equation}
\rho=6\pi G\left(\dot{H}^{-1}HvV_{SQ}\right)
\end{equation}
\begin{equation}
p=-\left(1+\frac{2\dot{H}}{3H^2}\right)\rho=w\rho ,
\end{equation}
in which $H=\dot{a}/a$, with $a$ the scale factor of the flat
universe. Now, from the Friedmann equation $H^2=8\pi G\rho/3$
derived from our Lagragian density, it follows that
\begin{equation}
\dot{H}=\pm 4\pi GvV_{SQ} .
\end{equation}
In addition, the equation of motion for coordinate $q$ has the
general form
\begin{equation}
v\dot{v}=-(1-v^2)F\left(H,v,m_0,V_{SQ}\right) .
\end{equation}
It can be shown that the function $F$ is always divergent provided
$v^2\neq 1$. Thus, in order to ensure regularity of the whole
model we should require that $v^2=1$ which is just the remaining
necessary condition for making the present model self-consistent;
i.e. $\phi^2=v^2=1$. In this way, we have for the Hubble function
\begin{equation}
H=H_0\pm 4\pi GV_{SQ}t ,
\end{equation}
with $H_0$ an integration constant playing the role of a
cosmological constant. We have then the solutions
\begin{equation}
a_{\pm}=a_0 e^{H_0 t \pm 2\pi GV_{SQ}},
\end{equation}
in which $a_0$ is the initial value of the scale factor. In Fig. 1
we give the evolution of the scale factor corresponding to these
solutions, as compared with that for a pure de Sitter universe.
Solution $a_-$ describes a universe which initially accelerates
with $w>-1$, then decelerates for a while to finally contract all
the way down to zero. Such a solution can be seen to violate the
second law of thermodynamics and therefore will not be here
considered as a realistic solution. Moreover, present estimates of
the parameter $w$ seem to place its value slightly beyond the de
Sitter barrier, a case which can never be described by solution
$a_-$. Solution $a_+$ has not these shortcomings. It corresponds
to what has been denoted as a phantom universe \cite{Gonza:2004}
characterized by a parameter $w<-1$ and will be taken in this
report as the physical solution representing the current evolution
of the universe.

\begin{figure}
  \includegraphics[height=.3\textheight]{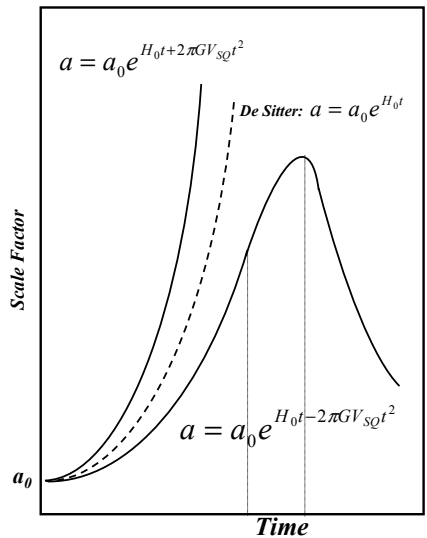}
  \caption{Time evolution of the sub-quantum cosmic solutions as compared to
that of de Sitter universe. Solution $a_-$ corresponds to an
equation of state with $w>-1$ and violates the second law of
thermodynamics; solution $a_+$ corresponds to a phantom model with
equation of state $w<-1$. Problems in such a model are benigner
than in known phantom models as it is stable, does not show any
singularities at finite time in the future, corresponds to a
positive kinetic term and violates dominant energy condition only
quantum-mechanically.}
\end{figure}

The point now is that the Lagragian density and both the energy
density and pressure become all zero, while the universe reduces
to a de Sitter universe, in the classical limit where the
sub-quantum potential vanishes. Thus, the main assumption of the
present model is to interpret that it is the sub-quantum effect
originated by the radiation particles that constitutes the cosmic
microwave background that is the unique cause making the universe
to accelerate. If so, we had accomplished a most economical model
justifying the current acceleration of the universe without
introducing any ac hoc mysterious dark energy field or modifying
the Hilbert-Einstein gravity.

\section{A benigner phantom universe}

Solution $a_+$ actually describes what we can call a benign
phantom universe. In fact, even though it corresponds to a
tracking equation of state with $w<-1$, but very close to -1 for
most of its evolution and the energy density is an increasing
function of the cosmological time \cite{Gonza:2004}, it can be
associated with a stable field theory having a kinetic term
$\dot{\phi}^2=\dot{q}^2 > 0$, shows no future singularity of the
big rip kind and violates the dominant energy condition only
quantum-mechanically, i.e. we always have $p+\rho=-V_{SQ}$ which
is a permissible violation of such a condition. We finally note
that our model makes it compatible the current dominance of a
sub-quantum energy phase with the previous matter domination in
the universe. The ultimate reason for this consists in the
realization that in a sub-quantum description all matter fields
entering the Lagrangian considered by Amendola, Quartin, Tsujiwara
and Waga \cite{Aqtw:2006} behave like though they were pure
radiation just at the coincidence time.

\section{Conclusion}

The main conclusion from this report is that the current
acceleration of the universe should be described by a benigner
phantom model which does not contain any extra fluid or
modification of gravity but just the quantum effects associated
with the existence of a sub-quantum potential for the CMB,
superposed or not to a cosmological constant.

\acknowledgements

\noindent This work was supported by MEC under Research Project
No. FIS2005-01181. The author benefited from discussions with C.
Sig\"{u}enza, A. Rozas, S. Robles, and J.A. Jim\'{e}nez Madrid.

\end{document}